\documentclass{INTERSPEECH2023}

\interspeechcameraready

\title{MaskedSpeech: Context-aware Speech Synthesis with Masking Strategy}
\name{Ya-Jie Zhang\textsuperscript{1}, Wei Song\textsuperscript{1}, Yanghao Yue, Zhengchen Zhang, Youzheng Wu, Xiaodong He \thanks{\textsuperscript{1}  Equal contributions. \\This work was supported by the National Key R\&D Program of China under Grant No. 2020AAA0108600.}}

\address{JD Technology Group}
\email{\{zhangyajie23,songwei11,yueyanghao,zhangzhengchen1,wuyouzheng1,hexiaodong\}@jd.com}

\begin{document}

\maketitle
 
\begin{abstract}

Many speech synthesis systems only consider the information within each sentence and ignore the contextual semantic and acoustic features.
This makes it inadequate to generate high-expressiveness paragraph-level speech.
In this paper, a context-aware speech synthesis system named MaskedSpeech is proposed, which considers both contextual semantic and acoustic features.
Inspired by the masking strategy in speech editing research, 
the acoustic features of the current sentence are masked out and concatenated with those of contextual speech, and further used as additional model input. 
Furthermore, cross-utterance coarse-grained and fine-grained semantic features are employed to improve the prosody generation. 
The model is trained to reconstruct the masked acoustic features with the augmentation of both the contextual semantic and acoustic features.
Experimental results demonstrate that the MaskedSpeech outperformed the baseline systems significantly in terms of naturalness and expressiveness.

\end{abstract}

\section{Introduction}
\label{sec:intro}
With the development of deep learning technology, current text-to-speech (TTS) systems~\cite{tacotron2,fastspeech,durian,dian} could generate high-quality speech.
However, most speech synthesis systems only consider the information in the current sentence to be synthesized and ignore the contextual text and speech. 
With the rising requirements on the naturalness and expressiveness of TTS systems, such as synthesizing audiobooks, these speech synthesis systems lack the ability to generate paragraph-level speech with adequate naturalness and expressiveness, as human speech is context coherent in a paragraph.

Thus, recent works~\cite{pse,contextaware,zyjbert,oplustil2021comparing,xiao2020improving,xue2022paratts} incorporated semantic features of the current sentence or contextual sentences to improve the prosody of synthesized speech, especially for paragraph speech generation.
Xiao~\textit{et al}.~\cite{xiao2020improving} extracted semantic information from the current sentence by a pretrained BERT~\cite{bert} model and showed that the incorporation of semantic features could improve the prosody of synthesized speech. 
Then works~\cite{pse,contextaware,zyjbert,xue2022paratts} further employed cross utterances semantic features in the speech synthesis systems, and all of them showed significant rhythm-enhancing effectiveness.
Although contextual semantic information could improve the prosodic performance to a certain extent, these works do not take into account the variation of contextual acoustic features, which is also very important for generating expressive and natural speech for a paragraph.

Some works~\cite{contextaware,gst,vae} used acoustics features to improve the prosody modeling, but most of them extracted sentence-level prosody representation and lost the fine-grained information of contextual acoustic features.
In addition, for the reference-based models~\cite{gst,vae}, it's non-trivial to select proper reference audio.
~\cite{liu2021delightfultts} and~\cite{chen2021adaspeech} introduced an acoustic encoder to learn fine-grained prosody features instead of utterance-level features and claimed that their models could improve prosody generation and relive the one-to-many mapping issue.
However,~\cite{liu2021delightfultts} and~\cite{chen2021adaspeech} only consider speech from the current sentence and the prosody features need to be predicted from text when reference audio is unavailable during inference, this led to another one-to-many mapping issue, as the same text can potentially be used to predict acoustic features from different speech.

To improve prosody generation for a paragraph, we proposed MaskedSpeech in this paper, which takes FastSpeech2~\cite{fastspeech2} as network backbone and incorporates both contextual semantic and acoustic features.
The main contributions of MaskedSpeech contains:
\begin{itemize}
    \item Inspired by the speech editing works~\cite{a3t,wang2022context,tan2021editspeech}, we concatenate the acoustic features of previous and current sentences and mask out those from the current sentence, then those features are used as additional decoder input for the proposed network. The network could learn fine-grained prosody features from the contextual speech at the decoder part, and no prosody prediction from text is required.
    \item Phonemes of both previous and current sentences are concatenated and sent into the phoneme encoder of the proposed model, which helps the model learn fine-grained semantic features from the previous sentence.
    \item Furthermore, a cross-utterance (CU) encoder is used to extract coarse-grained sentence-level semantic representation from neighboring sentences, this allows the proposed model to capture semantic correlation between contextual sentences.
\end{itemize}

The proposed MaskedSpeech is trained by reconstructing the masked acoustic features, by utilizing the contextual text, contextual speech, and sentence-level semantic representation, which makes MaskedSpeech able to improve prosody modeling and alleviate the one-to-many mapping issue naturally.

\section{Related work}
\label{sec:realted_work}

Unsupervised speech representation learning~\cite{wav2vec2,wav2vec2xlsr,hubert} has demonstrated its superiority in the field of speech recognition and related downstream tasks. 
Although the success of speech representation learning in speech understanding tasks, it's not suitable for the speech synthesis task due to its training strategy.
MAM~\cite{chen2020mam} and FAT-MLM~\cite{zheng2021fused} masked several acoustic segments randomly and used the masked acoustic features as network input, then Transformer-based~\cite{vaswani2017attention} network is employed to reconstruct the masked regions. 
But the quality of the reconstructed acoustic features could not meet the requirement of speech synthesis.
Bai~\textit{et al}.~\cite{a3t} proposed a network that uses masked acoustic features as input and the network is trained to predict high-quality acoustic features, then their proposed network could be further applied to speech synthesis and speech editing tasks.
CampNet~\cite{wang2022context} introduced a speech editing model, the acoustic features are randomly masked out by a prior ratio and then sent to the decoder for masked speech reconstruction. 
Each decoder layer used an attention mechanism to learn contextual features from unmasked part.

All the mentioned works for speech reconstruction try to predict high-quality and contextual consistent acoustic features by learning the context relationship of masked and unmasked speech.
Inspired by these works, we incorporate the masking strategy into cross-utterance speech synthesis task to improve the prosody modeling, with the assumption that both the textual and acoustic context could facilitate the prosody improvement of speech synthesis.
Instead of masking on word-level speech region for speech editing tasks, the acoustic features of the current sentence are concatenated with those of previous sentence, and then acoustic features from the current sentence are masked out.
In other words, we proposed a novel context-aware speech synthesis method that reconstructs speech for the current sentence with improved prosody by utilizing both the semantic and acoustic features of contextual sentences.

\section{Proposed Method}
\label{sec:masked_TTS}
The proposed MaskedSpeech utilizes FastSpeech2~\cite{fastspeech2} as the network backbone, which further fuses contextual semantic and acoustic features to improve prosody generation.
MelGAN~\cite{kumar2019melgan} is used to convert acoustic features to speech in the experiments.


\subsection{The MaskedSpeech Model}

\begin{figure}[t]
\centering
\includegraphics[width=7.7cm]
{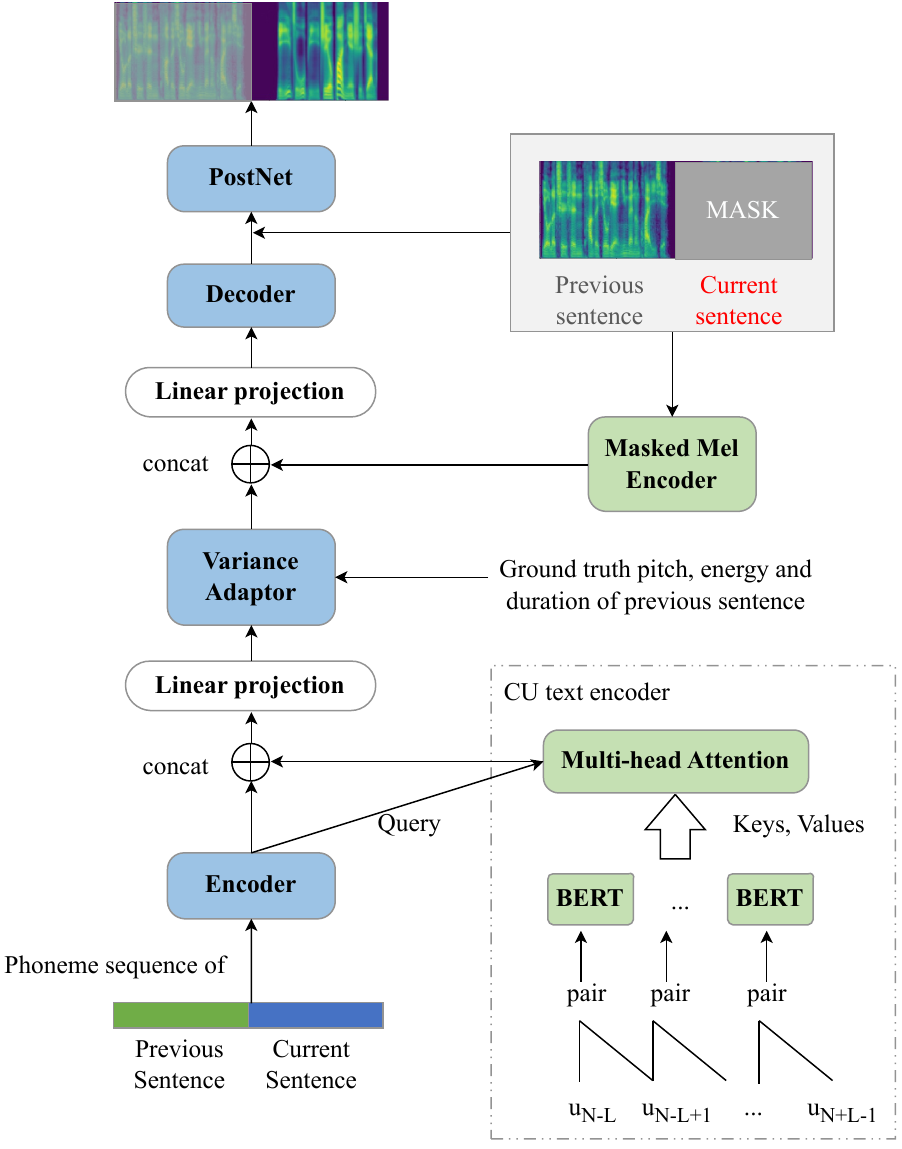}
\vspace{-1mm}
\caption{\label{fig:FIG1} The architecture of the MaskedSpeech model.}
\vspace{-5mm}
\end{figure}

Fig.~\ref{fig:FIG1} illustrates the network architecture of the proposed MaskedSpeech model. 
The backbone network consists of a phoneme encoder, a variance adaptor, a decoder, and a PostNet module.
To improve prosody for the speech in a paragraph by contextual features, concatenated sentences are used as the input of the phoneme encoder, a CU text encoder is employed to obtain CU semantic features from neighboring sentences. Furthermore, a masked mel-encoder is designed to extract local prosody features from the contextual speech, and then the decoder could learn global prosody dependency from the contextual speech.
By conditioning on contextual semantic and acoustic features, the proposed MaskedSpeech alleviates the one-to-many mapping issue and could generate speech with improved prosody for a paragraph, and the generated speech is more coherent and consistent with contextual speech.

\subsubsection{Learning from contextual text}

The prosody of the current sentence could be impacted by the relative sentence position, discourse relations (DRs) in neighboring sentences, the emotion of contextual sentences, etc. To improve the prosody generation by contextual text, we incorporate both fine-grained contextual phonemes and coarse-grained sentence-level semantic features into the proposed model.

The phoneme sequences of the previous sentence and current sentence are concatenated as the input of the phoneme encoder, rather than only using the phonemes of the current sentence in general methods.
Thus, the phoneme encoder could learn phoneme representation which takes previous sentence into consideration implicitly, and then the variance adaptor could generate more expressive prosody features.

Further, a CU encoder is used to incorporate sentence-level semantic representation.
Following the work~\cite{pse}, paired BERT embedding (PBE) is used to extract sentence-level coarse-grained semantic features between sentence pairs in the CU encoder.
As shown in Fig.~\ref{fig:FIG1}, the neighboring sentences $\{u_{N-L},...,u_N,u_{N+1},...,u_{N+L-1}\}$ can be used to derive 2$L$ sentence pairs, where $u_N$ means the current sentence and $L$ is the number of sentences taken into account in the past or following context.
Then PBE is extracted by a pre-trained BERT model from the $[CLS]$ token of a sentence pair.
A multi-head attention module is used to incorporate PBEs into the proposed model, where PBEs are taken as keys and values, and the output of the phoneme encoder is considered as the query.
The output of the CU encoder is concatenated with the output of the phoneme encoder, then linear projected.

\subsubsection{Learning from contextual speech}
Human speech is naturally coherent in a paragraph, the prosody and emotion of the current sentence could be affected by the contextual speech.
Contextual speech usually contains rich, expressive, and detailed prosody information, which could be utilized to improve the prosodic performance of the current sentence in speech synthesis systems, especially for the speech generation in a paragraph. 

A masked mel-encoder is incorporated into the FastSpeech2 network in the proposed MaskedSpeech. 
The acoustic features from contextual sentences, 
\textit{i.e.}
the mel-spectrograms of the previous and current sentences are concatenated and those from the current sentence are masked with a special~\textit{MASK} token, then those features are taken as the input of the masked mel-encoder.
The masked mel-encoder consists of two 1-dimensional convolutional (Conv1d) layers, it could learn to extract local prosody features from the acoustic features in the previous sentence.
ReLU activation function is used in each Conv1d layer, followed by layer normalization~\cite{ba2016layer} and dropout~\cite{srivastava2014dropout}.

The output of the masked mel-encoder is concatenated with the output of the variance adaptor and then projected to the dimension of the decoder. 
The decoder consists of several self-attention-based Conformer~\cite{gulati2020conformer} blocks, so the decoder could learn both local and global prosody dependency implicitly from the contextual acoustic features. 
As shown in Fig.~\ref{fig:FIG1}, the predicted acoustic features of the previous sentence are replaced with ground truth features at the decoder's output part, then a Post-Net module with 5 Conv1d layers is used to refine the predicted acoustic features. 
As the PostNet consists of Conv1d layers and learns local features, it could improve the boundary prosody transition for the current sentence.

\subsection{Training and Inference Strategies}
\label{sec:inference}

\subsubsection{Training}
During training, the model is trained to reconstruct the masked acoustic features and predict the prosody features of the current sentence.
In other words, we only compute training objectives for the current sentence. The losses include phoneme-level duration prediction loss, pitch and energy prediction losses, and mel-spectrogram reconstruction loss.
Mean squared error (MSE) is used as the training objectives for pitch, energy, and duration prediction, and mean absolute error (MAE) is used for the mel-spectrogram reconstruction. 

\subsubsection{Inference}
\label{sec:inference}
    
Different from the training procedure, where ground truth contextual text and speech are available and consecutive with the current sentence, there is no ground truth previous speech in the inference procedure, although the contextual text is available for a paragraph speech synthesis. 
To solve this discrepancy, we tried to select text-audio pair randomly from the training corpus and use it as the contextual speech and phonemes for the proposed model. As this random selection would make a difference compared to the training strategy, an AB preference test is conducted to compare the system performance with ground truth contextual speech and randomly selected contextual speech, the AB test result showed that no significant difference was observed for these two different kinds of contextual speech.
We could also select contextual speech with the expected style and prosody deliberately, to generate speech with the desired prosody.
In addition, 
when a single sentence and corresponding masked speech~(\textit{i.e.} a few words' acoustic features are masked out) are used as the network input, the proposed MaskedSpeech turns into a speech editing model, which can achieve inserting, replacing, and deleting part of audio by editing the corresponding text, as the model is trained to recover the masked acoustic region given the contextual text and speech.

\section{Experiments}
\label{sec:experiments}

\begin{table}[t]
\caption{The naturalness MOS results of different models with 95\% confidence intervals.}
\label{tab:mos}
 \renewcommand\arraystretch{1.2}
\centering
\renewcommand\tabcolsep{12.0pt}
\begin{tabular}{l c c}
\toprule
\textbf{Model} & \textbf{MOS} \\
 \hline
Ground truth & 4.43$\pm$0.07 \\
Fastspeech2  \cite{fastspeech2}   &4.05$\pm$0.08  \\
MaskedSpeech & \textbf{4.21$\pm$0.07} \\
\bottomrule
\end{tabular}
\vspace{-5mm}
\end{table}



\subsection{Experimental Setup}

A Chinese Mandarin audiobook corpus with 29 hours of recordings was used in the experiments, this corpus was recorded by a professional male speaker.
The text and recordings in this corpus are consecutive. 
200 sentences were selected for testing, and 200 sentences for validation, the remaining sentences were used for training.
The test set contains several paragraphs of consecutive sentences for evaluation with ground truth contextual text and speech.
The audio files were down-sampled to 16kHz sampleing rate.
80-dimensional mel-spectrograms were used as the acoustic features in the experiment and were extracted with a frame shift of 12 ms and frame length of 48 ms.
The phoneme boundaries and duration were extracted by Montreal forced alignment (MFA) \cite{mcauliffe2017montreal} tool. 
Pitch was extracted the PyWORLD\footnote{\url{https://github.com/JeremyCCHsu/Python-Wrapper-for-World-Vocoder}} toolkit.

The pre-trained ``bert-base-chinese'' model~\cite{wolf-etal-2020-transformers} released by Huggingface\footnote{\url{https://github.com/huggingface/transformers}} was used to extract cross utterance sentence embedding in the CU encoder. 
The BERT model consists of 12 Transformer layers, each layer contains 768 hidden units and 12-heads are used for each attention layer.
It is worth mentioning that the parameters of the BERT model are fixed during training.
The multi-head attention in the CU encoder has 4 heads and 512 hidden units.
In our experiments, PBEs are generated by 2 preceding sentences, 2 following sentences, and the current sentence. 
A filter size of 256 and kernel size of 3 is used in each Conv1d layer of the masked mel-encoder.

The proposed model is trained by Adam~\cite{kingma2014adam} optimizer, with a batch size of 16 and noam~\cite{vaswani2017attention} learning rate schedule.
The learning rate is warmed up to a maximum value of $10^{-3}$ in the first 4000 steps and then exponentially decayed.
Adam optimizer is configured with $\beta_{1}$=0.9, $\beta_{2}$=0.98 and $\epsilon$=$10^{-9}$. 
The model is trained by a total of 200,000 steps, with L2 parameter regularization of weight 1e-5.

\begin{table*}[t]
\centering
\caption{Subjective preference scores (\%) among different models, where N/P denotes ``No Preference" and $p$ means the $p$-value of $t$-test between two systems. }
\label{tab:psevspro}
\vspace{-6pt}
\begin{tabular}{c c c c c c c}
\toprule
\textbf{Fastspeech2} &  \textbf{PBE} & \textbf{MaskedSpeech w/o PBE} & \textbf{Random} &  \textbf{MaskedSpeech} & \textbf{N/P} & \textbf{p} \\
\midrule
19.5 & -  & -  & -&  \textbf{49.5}  &  31.0 &  \textless 0.001 \\
- & 23.0  & - & - & \textbf{54.0}   &   23.0 & \textless 0.001 \\
- & - & 26.5 & -& \textbf{38.0}   &  35.5 &  0.0425 \\
- & - & - & 27.5 &  29   &  43.5 &  0.779 \\

\bottomrule
\end{tabular}
\vspace{-5mm}
\end{table*}

\subsection{Subjective Evaluation}

To evaluate the quality of the proposed MaskedSpeech model, the Mean Opinion Score (MOS) evaluation is conducted to evaluate the naturalness and expressiveness of the synthesized speech\footnote{The audio samples can be found at \url{https://speech11.github.io/MaskedSpeech/}.}. 
A Fastspeech2 model was trained and used as the baseline model.
20 sentences were randomly selected from the test set and synthesized by different models, each audio is rated by 10 native Chinese listeners. 
The listeners are asked to rate the speech's naturalness and expressiveness on a scale from 1 to 5 with intervals of 0.5. 
Score 1 means the speech is unnatural completely and score 5 means the speech is completely natural.
For the proposed model, ground truth context speech and text are used for the audio generation.
Tab.~\ref{tab:mos} shows the MOS results.
Our proposed MaskedSpeech model achieved a higher MOS score than the baseline Fastspeech2 model in terms of naturalness and expressiveness.
This indicates that with the help of both contextual semantic and acoustic features, the proposed model could improve the prosody generation.





\subsection{Ablation Study on Contextual Features}
\label{sec:ablation}

In this section, ablation study is conducted to evaluate the effectiveness of both contextual semantic and acoustic features. 
The masking strategy is expected to help the synthesized audio to have a suitable and consecutive speaking style and prosody contour under the condition of the neighboring utterances.
In addition to masking strategy of acoustic features, there are some contributions corresponding to contextual semantic information which can further improve prosody modeling.
To evaluate the effectiveness of these contextual features, three AB preference tests were conducted for the ablation study.
First, we removed both CU encoder and masked mel-encoder from MaskedSpeech, which is a pure \textbf{Fastspeech2} model.
Then, we removed only masked mel-encoder, which is a \textbf{PBE}-based model.
In addition, we removed only the CU encoder, which is referred to~\textbf{MaskedSpeech w/o PBE} model (``w/o'' means without).
The MaskedSpeech model was compared with Fastspeech2 model, PBE-based model, and MaskedSpeech w/o PBE model separately.
Tab.~\ref{tab:psevspro} shows the results of AB preference tests.
For AB preference tests, the synthesized audio is concatenated with its previous recording to let the rater judge the consecutiveness and expressiveness of the speech.
As shown in Tab.~\ref{tab:psevspro}, MaskedSpeech outperforms the other three models significantly, which proves the effectiveness of using contextual semantic and acoustic features.
In addition, removing the masked mel-encoder results in a larger decline than removing the CU text encoder, this proves that fine-grained prosody features learned from the contextual speech play an important role in improving the prosody modeling.

As mentioned in section~\ref{sec:inference}, when the ground truth previous recording is not available, we randomly sample an audio from the corpus instead and use it as the contextual speech.
The method is referred to as~\textbf{Random}.
An AB preference test was conducted to evaluate it's performance compared with~\textbf{MaskedSpeech}. 
The result is shown in the fourth line of Tab.~\ref{tab:psevspro}. 
The MaskedSpeech model with ground truth previous speech is slightly better than~\textbf{Random} but the difference is not significant. 
In general, in some cases like narrations and normal-style dialogues, using random audio with a general style is comparable to using ground truth previous speech with the same style. 
However, we found that in some emotional dialogues, sometimes the speaker deliberately uses more distinctive prosody to align with the context speech.
In this case, using ground truth context speech will be better than randomly selected one without the desired style.
In conclusion, random sampling provides a substitutable way when there is no ground truth contextual speech.
Furthermore, compared with random sampling, manually selecting contextual speech will derive more controllable and expressive synthesized speech.  



\begin{figure}[t]
\centering
\includegraphics[width=6.5cm]{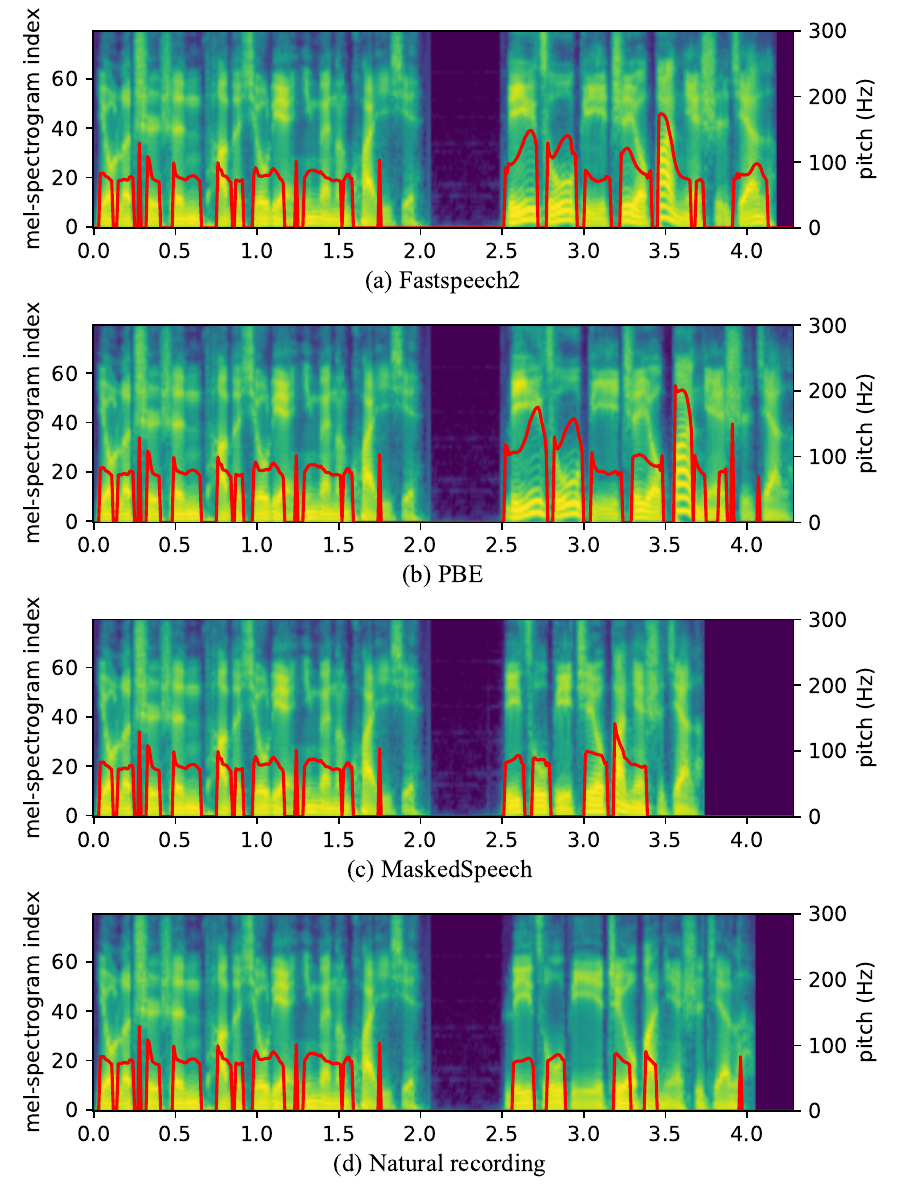}
\vspace{-2mm}
\caption{\label{fig:mels} A comparison between the mel-spectrograms of three models and natural recordings.}
\vspace{-5mm}
\end{figure}

\subsection{A Case Study}

Fig.~\ref{fig:mels} shows the mel-spectrograms of the speech synthesized by three different models and natural recording.
Each sub-figure shows the concatenated mel-spectrograms of previous and current sentences with distinct silence as the boundary.
The previous sentence is natural recording and the second sentence is the synthetic audio.
Fig.~\ref{fig:mels} (d) shows that the speaking speed of the nature recording is relatively faster and the fundamental frequency is lower.
MaskedSpeech perceives this property according to the contextual semantic and acoustic features and generates speech with a relatively faster speaking rate and lower pitch. 
This makes the prosodic transition of the consecutive sentences sound continuous and meaningful.

\section{Conclusion}
\label{sec:conclusion}

A context-aware speech synthesis system is proposed in this paper, with improved prosody generation by utilizing the contextual semantic and acoustic features from neighboring sentences.
The proposed model takes the concatenated and masked mel-spectrograms as augmented input and learns to reconstruct the masked mel-spectrograms. 
The phoneme encoder takes into account the phonemes from context sentences and those of the current sentence, and a CU encoder is used to extract cross-utterance semantic representation.
Experimental results showed the effectiveness of the proposed MaskedSpeech model.

\newpage
\bibliographystyle{IEEEtran}
\bibliography{mybib}

\end{document}